\documentclass[seceq]{ptptex}
\usepackage{graphicx}
\notypesetlogo                       
\title{
High-Temperature Expansion  
        of the Free Energy in the Two-Dimensional XY Model %
}

\author{
Hiroaki \textsc{ARISUE}%
}

\inst{
Osaka Prefectural College of Technology\\ 
Saiwai-cho, Neyagawa, Osaka 572-8572, Japan
}

\abst{
  We extend the high-temperature series of the free energy 
for the XY model in two dimensions
to order $\beta^{48}$ from the previous order of $\beta^{22}$
by applying an improved algorithm 
of the finite lattice method.
The long series obtained allows us to conclude that 
the behavior of the free energy is
consistent to high accuracy with what is expected when 
the phase transition of the model is of the Kosterlitz-Thouless type. 
}
\begin{document}
\maketitle
\section{Introduction}

We investigate the XY model in two dimensions.
It is believed to exhibit a phase transition 
of the  Kosterlitz-Thouless (K-T) type\cite{Kosterlitz},
driven by the condensation of vortices.
Through use of renormalization group arguments,
it was predicted that the correlation length 
possesses an essential singularity at the 
transition temperature $T_c$ of the form
\begin{equation}
   \xi \sim \exp\left[\frac{b}{t^{1/2}}\right] ,  \label{eqn:xi_cr}
\end{equation}
where $t=T/T_c-1$ is the reduced temperature and $b$ is a non-universal constant.
The free energy and its temperature derivatives
(i.e., the internal energy and the specific heat)
have also been predicted to behave as
\begin{equation}
   f \sim \exp\left[-\frac{2b}{t^{1/2}}\right] + [{\rm\ a\ regular\ term}] .
                                                   \label{eqn:f_cr}
\end{equation}
The first term on the right-hand side of Eq.~(\ref{eqn:f_cr}) has 
an essential singularity at $t=0$. 
This term itself and its derivatives 
are zero at the critical temperature.
The second term is regular at the critical temperature.

The behavior expressed in (\ref{eqn:xi_cr}) for the correlation length has been 
well established both by numerical simulations
and the high-temperature expansion.
The standard Monte Carlo simulation\cite{Edwards1991,Gupta1992} gives 
$b=2.15(10)$ with $\beta_c=1.130(15)$  
and $b=1.70(20)$ with $\beta_c=1.118(5)$ 
for the square lattice. 
(Here $\beta_c$ is the inverse critical temperature, which is defined below.)
The more precise values $b=1.800(2)$ with $\beta_c=1.1208(2)$ 
and $b=1.776(4)$ with $\beta_c=1.1199(1)$ were
obtained using the finite-size scaling technique\cite{Schultka1994} 
and the renormalization group finite-size scaling method\cite{Hasenbusch1997},
respectively.
In the latter approach, the renormalization group
flow of the observable was matched 
with that of the exactly solvable BCSOS model.
The latter value of $\beta_c$ was recently confirmed 
by large scale Monte Carlo simulations
on a $2048\times 2048$ lattice using the finite-size scaling method.
\cite{Hasenbusch2005}
A high-temperature expansion for the correlation length \cite{Butera1993,
Campostrini1996} 
gives the slightly smaller value $b=1.67(4)$ with $\beta_c=1.118(3)$.

In contrast to the situation for (\ref{eqn:xi_cr}), 
the behavior expressed by (\ref{eqn:f_cr}) for the free energy 
or its temperature derivatives has not yet been confirmed in 
numerical simulations or in high-temperature expansions.
The reason for this is the following.
The free energy and its derivatives are dominated by the regular term 
near the critical temperature, and
the signal of the singularity in the first term is relatively weak.
For this reason, a much more precise simulation is needed
to confirm the behavior given in (\ref{eqn:f_cr}) for these quantities 
than in the case of the correlation length.
For the high-temperature expansion also,
the presence of the regular term in (\ref{eqn:f_cr}) makes it difficult to
use standard analyzing methods of the series 
(such as the Pad\'{e} approximation 
and the inhomogeneous differential approximation) 
to pick out the singular term.
However, we believe that 
the contribution of the singular term 
to the coefficients in the high-temperature series
will dominate at sufficiently high orders. 
If this is indeed the case,
we could directly compare the high-temperature series of the free energy
with an expansion series of a form like that of the first term 
on the right-hand side of Eq.~(\ref{eqn:f_cr}) 
if we calculate a sufficiently long series for the free energy.
In this paper, we extend the previously derived high-temperature series
for the free energy\cite{Campostrini1996} 
to a series that is more than two times longer. 
This makes it possible to reveal the singular behavior 
of Eq.~(\ref{eqn:f_cr}). From an analysis of the obtained series
we obtain the value of $\beta_c=1.1176(33)$ for the inverse critical temperature
and $b=1.54(4)$ for the non-universal constant in Eq.(\ref{eqn:f_cr}). 

A commonly used method for series expansions 
is the graphical method\cite{Domb1974}.
However, in this method, one must list all the graphs that 
contribute to the desired order of the series.
An alternative, powerful method to generate the expansion series is 
the finite lattice method\cite{Enting1977,Arisue1984,Creutz1991}. 
It does not require that we list the graphs 
and it reduces the problem to that of calculating the partition functions for 
the relevant finite-size lattices. This is a rather straightforward procedure  
if we use the transfer matrix formulation.
In many cases, the finite lattice method generates longer series  
than the graphical method\cite{Bhanot1992,Guttmann1993,
Arisue1993,Arisue1994,Arisue1995,Bhanot1993,Guttmann1994b,
Arisue1999}.
Unfortunately, the original finite lattice method can generate 
a high-temperature series that is at most as long as that which can be obtained 
with the graphical method
in the case of the XY model in two dimensions. 

Here we apply an improved algorithm
of the finite lattice method developed by the author and Tabata
\cite{Arisue1995b,Arisue1995c} to generate a long series for the 
the XY model in two dimensions.
This improved algorithm is powerful in the case of models in which 
the spin variable at each site takes more than two values,
including the case that it takes an infinite number of values.
This algorithm was applied to generate a low-temperature series for  
the absolute value solid-on-solid (ASOS) model and 
high- and low-temperature series for the $q$-state Potts model 
in two dimensions. In both cases, it generates much longer series 
than the original finite lattice method.
The XY model in two dimensions can be mapped to a kind of 
solid-on-solid model, and the improved algorithm of the finite lattice method
enables us to obtain a series that is
two times longer than the series previously derived for the free energy.
We confirm from analysis of the obtained long series that
the free energy of the two-dimensional XY model behaves like Eq.~(\ref{eqn:f_cr}), 
with values of the critical temperature and the non-universal constant $b$ 
that are close to the values
obtained in studies of the correlation length.

\section{Algorithm}

 We consider the XY model defined on the square lattice. 
The Hamiltonian of this system is
\begin{equation}
      H = -\sum_{\langle i,j \rangle} J \vec{s}_i \vec{s}_j\;,
\end{equation}
where  $\vec{s}_i$ is a two-dimensional unit vector located at the lattice site $i$, 
and the summation is taken over all the pairs $\langle i,j \rangle$ 
of nearest neighbor sites.
The partition function at temperature $T$ is
\begin{equation}
      Z = \int \prod_{i} d\theta_i  \exp{\left( - \frac{H}{kT} \right) }\;,
\end{equation}
where $\theta_i$ is the angle variable of the spin 
$\vec{s_i}=(\cos{\theta_i},\sin{\theta_i})$.
 This model can be mapped exactly to a solid-on-solid model
whose partition function is
$$
Z = \sum_{\{ h\; |\, -\infty \le h_i \le +\infty  \}} 
\prod_{\langle i,j \rangle} I_{|h_i-h_j|}(\beta)\;,
$$
where $\beta=\frac{J}{kT}$, 
the quantities $I_n$ are modified Bessel functions,
the product is taken with respect to all the pairs of 
neighboring plaquettes,
and the variable $h_i$ at each plaquette $i$ 
takes an integer value between $-\infty$ and $+\infty$.

 The improved algorithm of the finite lattice method 
to generate the high-temperature expansion series 
for the free energy of this model employed here is the following.
 We first calculate the partition function for each of the
$l_x \times l_y$ finite-size rectangular lattices 
with a restricted range of the values of the plaquette variables $\{h\}$:
\begin{equation}
    Z(l_x,l_y;h_{+},h_{-})
       =  \sum_{\{h\; |\,  h_{-} \le h_i \le h_{+} \}} 
                  \prod_{\langle i,j \rangle} I_{| h_i - h_j |}(\beta) \;.
\end{equation}
Here, 
each plaquette variable $h_i$ is 
restricted as $ h_{-} \le h_i \le h_{+}$  
(where $h_{-} \le 0$ and $h_{+} \ge 0 $). 
We define the size of the finite lattice so that an  
$l_x \times l_y$ lattice involves $l_x \times l_y$ plaquettes,
including the bonds and sites on their boundary.
For instance, the $1 \times 1$ lattice consists of a single plaquette, 
including 4 bonds and 4 sites.
We take into account finite-size lattices with $l_x=0$ and/or $l_y= 0$.
An $l_x \times 0$ lattice consists of $l_x$ bonds and
$l_x +1$ sites with no plaquette.
The $0 \times 0$ lattice consists only of one site with no bond or plaquette.
 We take the boundary condition 
such that all the plaquette variables outside the $l_x \times l_y$ lattice 
are fixed to zero. 

 We then define $W$ of the $l_x \times l_y$ lattice and of the restricted range of
the plaquette variables recursively as
\begin{eqnarray}
\!\!\!\!\!\!\!\!\!\!\!\!\!
&& W(l_x, l_y;h_{+},h_{-})
        = -\log Z(l_x, l_y;h_{+},h_{-}) \nonumber\\
\!\!\!\!\!\!\!\!\!\!\!\!\!
&& \quad -  \sum_{\scriptstyle 
               0 \le l_x^{\prime} \le l_x,\ 
               0 \le l_y^{\prime} \le l_y,\ 
               0 \le h_{+}^{\prime} \le h_{+}, \ 
               h_{-} \le h_{-}^{\prime} \le 0, 
           \atop\scriptstyle 
               l_x^{\prime} \ne l_x,\  
               l_y^{\prime} \ne l_y \ 
               h_{+}^{\prime} \ne h_{+},\ 
               h_{-}^{\prime} \ne h_{-}    }
\!\!\! (l_x-l_x^{\prime}+1)(l_y-l_y^{\prime}+1) 
        W(l_x^{\prime}, l_y^{\prime}; h_{+}^{\prime},h_{-}^{\prime} ). \nonumber\\
&& 
\end{eqnarray}
The initial form of $W(l_x, l_y;h_{+},h_{-})$ is as follows:
\begin{eqnarray}
&&
\!\!\!\!\!\!\!\! 
Z(0, 0; h_{+}, h_{-} ) = 1  \ \ \
\mbox{and} \ \ \ 
W(0, 0; h_{+}, h_{-} ) = 0.  \\
&& 
\!\!\!\!\!\!\!\! 
Z(l_x, 0; h_{+}, h_{-} ) = Z(l_x, 0; 0, 0)={I_{0}(\beta)}^{l_x} \ \ \
\mbox{for}\ \ l_x \ge 1\ \ \ 
\mbox{and} \nonumber\\
&& 
\!\!\!\!\!\!
W(l_x, 0; h_{+}, h_{-}) = \left\{
\begin{array}{ll}
 - \log{I_{0}(\beta)}\  
    &  \mbox{for}\ l_x=1\ \mbox{and} \  h_{+}= h_{-}=0, \\ 
   0 \ 
    &\mbox{for}\ l_x\ge 2\ \mbox{or} \ 
         \ h_{+} \ge 1\ \mbox{or} \ h_{-}\le -1.
\end{array}
\right.  \\
&& 
\!\!\!\!\!\!\!\! 
Z(1, 1; h_{+}, h_{-}) = \sum_{h=h_{-}}^{h_{+}} I_{|h|}(\beta)^{4} \ \ 
\mbox{and}  \nonumber\\
&&
\!\!\!\!\!\!
W(1, 1; 0, 0) 
    = -\log Z(1,1;0,0) - 2 W(1,0;0,0) - 2 W(0,1;0,0) 
                                     -4W(0,0;0,0)\nonumber\\
&&
\qquad\qquad\ 
    = -\log I_{0}(\beta)^4 + 4 \log I_{0}(\beta) = 0. \nonumber\\
&&
\!\!\!\!\!\!
W(1, 1; 1, 0) = -\log Z(1,1;1,0) 
    -   W(1,1;0,0)
    - 2 W(1,0;1,0) - 2 W(1,0;0,0)\nonumber\\
&&
\qquad\qquad\ \ \ \ \ 
    - 2 W(0,1;1,0) - 2 W(0,1;0,0)
    - 4 W(0,0;1,0) - 4 W(0,0;0,0)\nonumber\\
&&
\qquad\qquad\ 
      = -\log \{I_{0}(\beta)^4+I_{1}(\beta)^4\} 
           +4 \log I_{0}(\beta) \nonumber\\
&&
\qquad\qquad\ 
      = -\log \left[ 1+\left\{
        \frac{I_{1}(\beta)}{I_{0}(\beta)} \right\}^4 \right]
      =  O (\beta^4) . \\
&&
\qquad\ 
\cdots \nonumber
\end{eqnarray} 
The free energy density $f_s$ per site in the infinite-volume limit is given by 
\begin{equation}
      f_s =  \sum_{ l_x,l_y,h_{+},h_{-}} 
                  W( l_x, l_y;h_{+},h_{-} ) . 
      \label{eqn:free_energy}
\end{equation}

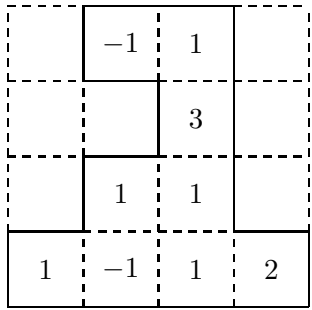
\begin{figure}[b!]
\setlength{\unitlength}{1.0mm}
\begin{picture}(100,50)
\put(40,0){\begin{picture}(100,50)(-5,0)
%
\multiput(5,5)(2,0){20}{\line(1,0){1}}
\multiput(5,15)(2,0){20}{\line(1,0){1}}
\multiput(5,25)(2,0){20}{\line(1,0){1}}
\multiput(5,35)(2,0){20}{\line(1,0){1}}
\multiput(5,45)(2,0){20}{\line(1,0){1}}
\multiput(5,5)(0,2){20}{\line(0,1){1}}
\multiput(15,5)(0,2){20}{\line(0,1){1}}
\multiput(25,5)(0,2){20}{\line(0,1){1}}
\multiput(35,5)(0,2){20}{\line(0,1){1}}
\multiput(45,5)(0,2){20}{\line(0,1){1}}
\put(5,5){\line(1,0){40}}
\put(5,15){\line(1,0){10}}
\put(35,15){\line(1,0){10}}
\put(15,25){\line(1,0){10}}
\put(15,35){\line(1,0){10}}
\put(15,45){\line(1,0){20}}
\put(5,5){\line(0,1){10}}
\put(15,15){\line(0,1){10}}
\put(15,35){\line(0,1){10}}
\put(25,25){\line(0,1){10}}
\put(35,15){\line(0,1){30}}
\put(45,5){\line(0,1){10}}
\put(10,10){\makebox(0,0){$1$}}
\put(20,10){\makebox(0,0){$-1$}}
\put(30,10){\makebox(0,0){$1$}}
\put(40,10){\makebox(0,0){$2$}}
\put(20,20){\makebox(0,0){$1$}}
\put(30,20){\makebox(0,0){$1$}}
\put(30,30){\makebox(0,0){$3$}}
\put(20,40){\makebox(0,0){$-1$}}
\put(30,40){\makebox(0,0){$1$}}
\end{picture}
}
\end{picture}
\caption{
An example of a polymer in the standard cluster expansion.
}
\label{figure:polymer1}
\end{figure}

 In the standard (graphical) cluster expansion of the free energy for this model, 
a cluster is composed of polymers, and 
each of the polymers consists of connected plaquettes. 
An example of a polymer can be seen in Fig.~1.
A value $h_i\ ( \neq 0 )$ is assigned to each site $i$ of the polymer. 
 We can assign to each cluster 
two numbers, $h_{\rm max}\ (\ge 0)$ and $h_{\rm min}\ (\le 0)$, 
which are the maximum and the minimum, respectively, of the quantities $h_i$
in all the sites of the polymers of which the cluster consists.
 Then, we can prove\cite{Arisue1984} 
that the Taylor expansion of $W(l_x, l_y;h_{+},h_{-})$ 
with respect to $\beta$ includes 
the contributions from all the clusters of polymers 
in the standard cluster expansion 
for which $h_{\rm max}=h_{+}$ and $h_{\rm min}=h_{-}$ 
and that can be embedded into the $l_x \times l_y$ lattice 
but cannot be embedded into any of its rectangular sub-lattices.
 The series expansion of $W(l_x, l_y;h_{+},h_{-})$ 
begins at order $\beta^{\, n(l_x,l_y,h_{+},h_{-})}$, with 
\begin{equation}
  n(l_x,l_y,h_{+},h_{-}) = \left\{
\begin{array}{ll}
   2(l_x + l_y) + 4(h_{+}+|h_{-}|)-6
                  & \mbox{for}\  h_{+}> 0 \ \mbox{and}\  h_{-}<0, \\
   2(l_x + l_y) + 4(h_{+}+|h_{-}|)-4 
                  & \mbox{for}\  h_{+}=0 \ \mbox{or} \ h_{-}=0.
                                            \label{eqn:order_to_start}
\end{array}
\right.
\end{equation}
An example of a cluster consisting of a single polymer 
that contributes to the lowest-order 
term of $W(l_x, l_y;h_{+},h_{-})$ is given in Fig.~2.
 Hence, to obtain the expansion series to order $\beta^N$,
we have only to take into account all combinations of the rectangular lattice
size $(l_x, l_y)$ and the range of the plaquette variables $(h_{+},h_{-})$ 
that satisfy the relation 
$ n(l_x,l_y,h_{+},h_{-}) \le N $
in the summation of Eq.~(\ref{eqn:free_energy}) 
and to evaluate each of the $W(l_x,l_y,h_{+},h_{-})$ to order $\beta^N$. 

\begin{figure}[ht!]
\setlength{\unitlength}{1.0mm}
\begin{picture}(100,50)
\put(40,0){\begin{picture}(100,50)(-5,0)
%
\multiput(5,5)(2,0){20}{\line(1,0){1}}
\multiput(5,15)(2,0){20}{\line(1,0){1}}
\multiput(5,25)(2,0){20}{\line(1,0){1}}
\multiput(5,35)(2,0){20}{\line(1,0){1}}
\multiput(5,45)(2,0){20}{\line(1,0){1}}
\multiput(5,5)(0,2){20}{\line(0,1){1}}
\multiput(15,5)(0,2){20}{\line(0,1){1}}
\multiput(25,5)(0,2){20}{\line(0,1){1}}
\multiput(35,5)(0,2){20}{\line(0,1){1}}
\multiput(45,5)(0,2){20}{\line(0,1){1}}
\put(5,5){\line(1,0){40}}
\put(5,15){\line(1,0){20}}
\put(35,15){\line(1,0){10}}
\put(25,45){\line(1,0){10}}
\put(5,5){\line(0,1){10}}
\put(25,15){\line(0,1){30}}
\put(35,15){\line(0,1){30}}
\put(45,5){\line(0,1){10}}
\put(10,10){\makebox(0,0){$h_{-}$}}
\put(20,10){\makebox(0,0){$-1$}}
\put(30,10){\makebox(0,0){$1$}}
\put(40,10){\makebox(0,0){$1$}}
\put(30,20){\makebox(0,0){$1$}}
\put(30,30){\makebox(0,0){$h_{+}$}}
\put(30,40){\makebox(0,0){$1$}}
\end{picture}
}
\end{picture}
\caption{
An example of a cluster consisting of a single polymer
that contributes to the lowest-order 
term of $W(l_x, l_y;h_{+},h_{-})$ with $l_x=l_y=4$, $h_{-}<0$ and $h_{+}>0$.
}
\label{figure:polymer2}
\end{figure}

Most of the CPU time required with this algorithm is used 
for the calculation of the relevant partition functions 
$Z(l_x, l_y;h_{+},h_{-})$ to order $\beta^N$.
These partition functions are calculated 
with the transfer matrix method using a procedure in which 
a finite-size lattice is built one plaquette at a time\cite{Enting1980,Bhanot1990}.
 The necessary CPU time and memory are proportional 
to $ N l_x l_y \ (h_{+}+|h_{-}|+1)^{\min{(l_x,l_y)}} $
and $ N \ (h_{+}+|h_{-}|+1)^{\min{(l_x,l_y)}}$,
respectively. 

If we had applied the original finite lattice method
straightforwardly, without introducing $h_+$ and $h_-$,
we would have had to calculate the partition function
while taking into account all the configurations with each of the 
plaquette variables ranging from $-N/4$ to $N/4$ for any 
necessary size of the lattice.
However, because we have introduced $h_+$ and $h_-$, we only need to
take into account the configurations 
with each of the 
plaquette variable ranging between $h_{-}$ and $h_{+}$,  
with $|h_{-}|+h_{+} \le (N-2(l_x+l_y))/4$.
This substantially reduces the number of the configurations 
that have to be taken into account, 
especially for larger lattices.
In fact, if we had used the original method, 
we would have been able to generate a high-temperature
series only to order $\beta^{24}$, instead of $\beta^{48}$.

\section{Series}
\begin{table}[b!]
\caption{High-temperature series coefficients for the free energy per bond 
of the XY model on the square lattice.}
\begin{center}
\begin{tabular}{r|l}
\hline
\hline
    $n$  &  \hspace{60mm}    $a_n$        \\
\hline
$   2 $ & \hspace{65.5mm} $                                             -1 $                                                 \\
$   4 $ & \hspace{63mm} $                                             -3 $  /  $                                       4 $ \\
$   6 $ & \hspace{63mm} $                                             -1 $  /  $                                       9 $ \\
$   8 $ & \hspace{64mm} $                                             31 $  /  $                                     192 $ \\
$  10 $ & \hspace{62mm} $                                            731 $  /  $                                     600 $ \\
$  12 $ & \hspace{59mm} $                                          29239 $  /  $                                    8640 $ \\
$  14 $ & \hspace{57mm} $                                         265427 $  /  $                                   35280 $ \\
$  16 $ & \hspace{54mm} $                                       75180487 $  /  $                                 5160960 $ \\
$  18 $ & \hspace{51mm} $                                     6506950039 $  /  $                               235146240 $ \\
$  20 $ & \hspace{46mm} $                                  1102473407093 $  /  $                             26127360000 $ \\
$  22 $ & \hspace{46mm} $                                  6986191770643 $  /  $                            158070528000 $ \\
$  24 $ & \hspace{39mm} $                              -1657033646428733 $  /  $                          49662885888000 $ \\
$  26 $ & \hspace{35.5mm} $                          -132067307951625029 $  /  $                         349709488128000 $ \\
$  28 $ & \hspace{31mm} $                         -123922162516396049161 $  /  $                       73815602724864000 $ \\
$  30 $ & \hspace{29mm} $                        -3988676736069063737749 $  /  $                      684416646144000000 $ \\
$  32 $ & \hspace{21mm} $                   -171797935005592667747929271 $  /  $                  9718351353033523200000 $ \\
$  34 $ & \hspace{19mm} $                  -4447072728482029389525377383 $  /  $                 87768860657084006400000 $ \\
$  36 $ & \hspace{14.4mm} $             -1419869374002650746096436008343 $  /  $              10036627359845371084800000 $ \\
$  38 $ & \hspace{13mm} $              -37739166891866120573923814513909 $ 
                                                                            /  $             100645068802893860044800000 $ \\
$  40 $ & \hspace{8mm} $            -79972307856390845424075199180330349 $  
                                                                            /  $           84753742149805355827200000000 $ \\
$  42 $ & \hspace{5mm} $          -3529608673175063035175810323068581713 $  
                                                                            /  $         1509431550094898846760960000000 $ \\
$  44 $ &              $      -55921787617500176735150797125026274875543 $  
                                                                            /  $      9949597756682679903049482240000000 $ \\
$  46 $ &               $   -1470445398138160496551515879389714706601279 $  \\
                                                            & \hspace{65mm} /  $    119621300301934947016208547840000000 $    \\
$  48 $ &               $-7033812372580933320904163849616698204389178239 $   \\
                                                            & \hspace{60mm}  / $ 287590409630251949139489907015680000000 $      \\
\hline
\end{tabular}             
\end{center}
\end{table}

 We have calculated the high-temperature expansion series 
for the free energy of the XY model on the square lattice 
to order $\beta^{48}$. 
 The obtained coefficients for the free energy $f_b$ per bond are listed 
in Table~I, 
where the quantities $a_n$ are defined through the relation
\begin{equation}
  f_b = \frac{f_s}{2}=  \sum_{n=1}^{N} a_n \left(\frac{\beta}{2}\right)^n.  \label{eqfe}
\end{equation}
 We have confirmed that each $W(l_x, l_y;h_{+},h_{-})$
in Eq.~(\ref{eqn:free_energy}) starts from the correct order in $\beta$, 
as given by Eq.~(\ref{eqn:order_to_start}).
 The first eleven terms of the series (to order $\beta^{22}$) coincide 
with those obtained by Campostrini et al.,\cite{Campostrini1996}
and we have generated thirteen new terms.

To obtain the long numerators and denominators appearing in Table I precisely,
we have used the following technique.
The Taylor series of the Bessel function $I_{n}(\beta)$ can be written
\begin{equation}
   I_{n}(\beta)= \sum_i \frac{b_{n,i}}{i!} \left(\frac{\beta}{2}\right)^i\;,
                                         \label{eqn:Bessel}
\end{equation}
where the coefficients $b_{n,i}$ are all integers.
The product of two Bessel functions can also be written as
\begin{equation}
   I_{n_1}(\beta)I_{n_2}(\beta)= \sum_i \frac{c_i}{i!} \left(\frac{\beta}{2}\right)^i\;,
                                         \label{eqn:product_Bessel}
\end{equation}
with
\begin{equation}
   {c}_i= \sum_{i'=0}^{i} \frac{i!}{i'!(i-i')!} b_{n_1,i'} b_{n_2,i-i'}\;.
\end{equation}
We note that the combinatorial factor $\frac{i!}{i'!(i-i')!} $  
is an integer and that $c_i$ is also an integer.
Thus, if we treat 
coefficients like $b_{n,i}$ and ${c}_i$ 
in Eqs.~(\ref{eqn:Bessel}) and (\ref{eqn:product_Bessel})
instead of the coefficients of the terms in the Taylor expansion,
then we need only treat integer values
throughout the calculation.
For the products and summations of the long integers that arise in these calculations,
we used the Chinese theorem of modulus\cite{Guttmann1993}.

The calculations were carried out on a workstation
at the Information Processing Center at OPCT and
on an Altix3700 BX2 at YITP of Kyoto University.

\section{Series analysis}

If the phase transition of the model is of K-T type, the free energy is
expected to behave as
\begin{equation}
     f(\beta) = A \beta^2 \exp\left[-\frac{2b}{(1-\beta / \beta_c)^{1/2}}\right] 
                +B(\beta) ,               \label{eqn:f_cr2}
\end{equation}
where 
$\beta_c$ is the inverse critical temperature and 
$b\ (>0)$ is the non-universal constant that appears 
in Eqs.~(\ref{eqn:xi_cr}) and (\ref{eqn:f_cr}).

 First, it is to be noted that the coefficients of the free energy in Table~I
exhibit repeated changes of sign. 
This is one indication that the phase transition in this model 
is of the Kosterlitz-Thouless type.
This property of the signs of the coefficients in the expansion series 
of the free energy for a model that undergoes a K-T type phase transition 
was first pointed out by Hasenbusch et al.\cite{Hasenbusch1994}
The coefficients in the Taylor expansion of the first term  
on the right-hand side of Eq.~(\ref{eqn:f_cr2}) with respect to $\beta$ 
exhibits repeated changes of sign in general,
with the order of the sign change depending on the values of $c$ and $\beta_c$, 
while the coefficients of the terms in the Taylor expansion 
for a quantity with a power law singularity do not exhibit changes of sign.
In fact, changes in sign of the coefficients is observed 
in the low-temperature series for the surface free energy of the Ising model
(i.e., the strong coupling series for the string tension 
of the $Z_2$ lattice gauge theory) 
in three dimensions\cite{Arisue1993}
and also in the low-temperature series of the free energy
for the ASOS model in two dimensions\cite{Hasenbusch1994,Arisue1995b},
both of which undergo a roughening phase transition
of the K-T type.

It is expected that
the coefficients of the terms in the series expansion of the free energy 
at sufficiently high orders will be dominated 
by the contribution from the Taylor expansion 
of the first singular term in Eq.~(\ref{eqn:f_cr2}).
With this in mind, rewriting the high-temperature series as 
\begin{equation}
   f = \sum_{n=1}^{N/2}   \tilde{a}_n x^n  
         \qquad \left(x 
                   =\left(\frac{\beta}{\beta_c}\right)^2\right) \;,
\end{equation}
we fit the K-T type function 
\begin{equation}
   f^{\mbox{\tiny (K-T)}}(x) 
  = A\, x \exp\left[-\frac{c_0 + c_1 x}{(1-x)^{\sigma}}\right]
  = \sum_n \tilde{a}^{\mbox{\tiny (K-T)}}_n x^n   
\end{equation}
to the high-temperature series under the condition that 
\begin{equation}
   \frac{\sum_{n=n_0}^{N/2} ( \tilde{a}^{\mbox{\tiny (K-T)}}_n - \tilde{a}_n )^2}
        {\sum_{n=n_0}^{N/2} {\tilde{a}_n}^2}     \label{eqn:criteria_param}
\end{equation}
be minimized. In the present case, we have $N/2=24$, and 
we have carried out the fitting for $n_0$ ranging from $7$ to $13$. 

\begin{figure}[b!]
\centerline{
\includegraphics[width=7cm,angle=270]{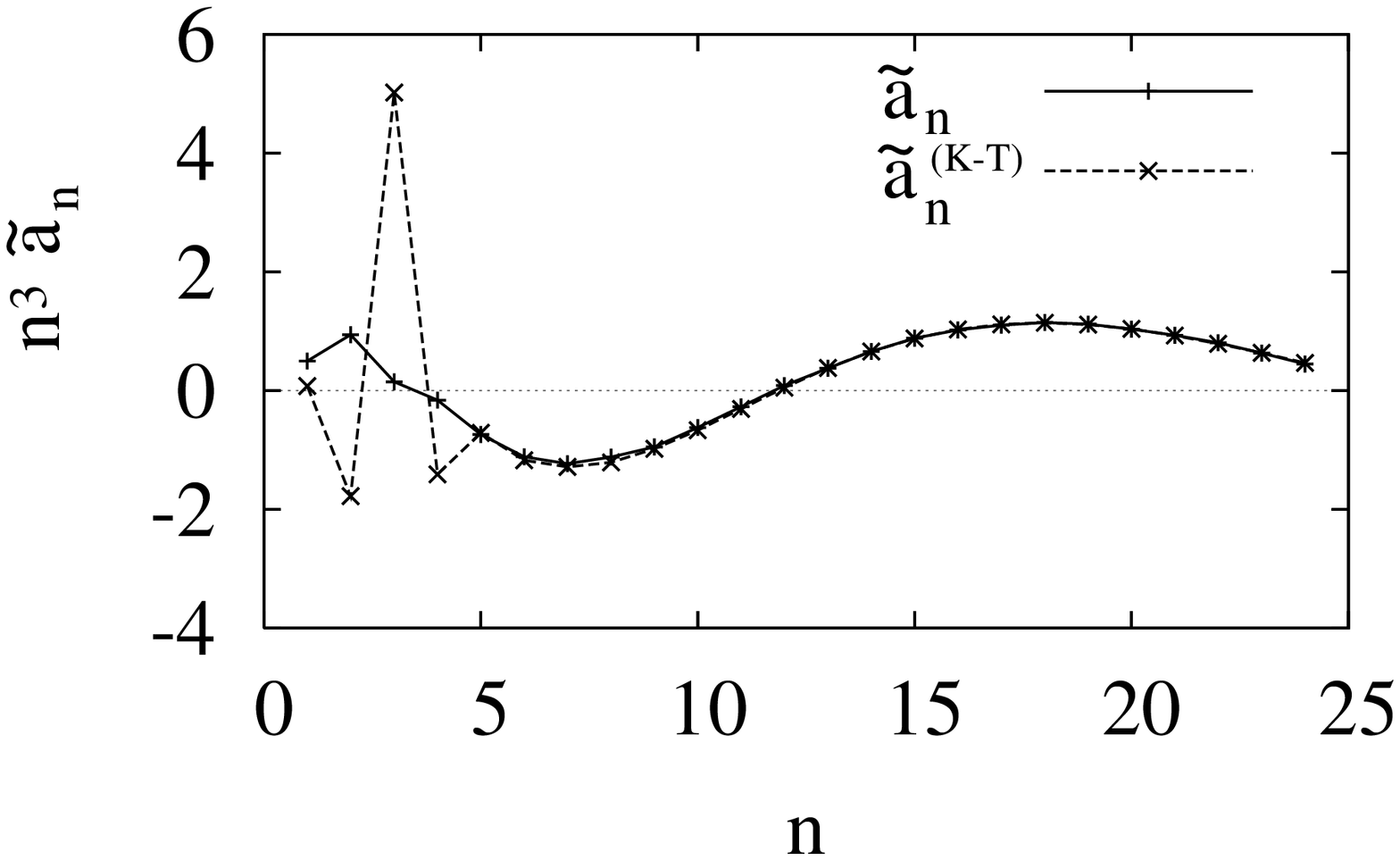}
}
\caption{The coefficients $\tilde{a}_n$ and 
$\tilde{a}^{\mbox{\tiny (K-T)}}_n$ multiplied by $n^3$.}
\end{figure}

We have performed three types of fittings. 
We first fixed the exponent $\sigma$ to $1/2$. 
In this case, the best fitting 
parameter values 
are $\beta_c=1.1176(33)$, $c_0=2.86(25)$ and $c_1=1.50(18)$.
The uncertainty on each value is due to  
the range of the values of $n_0$ adopted.
This best fitting value of the inverse critical temperature is consistent 
with the most precise value $\beta_c=1.1199(1)$, obtained
from the large-scale Monte Carlo simulation\cite{Hasenbusch2005},
and the value  $\beta_c=1.116(4)$ obtained from the high-temperature series 
for the correlation length and the magnetic susceptibility
\cite{Campostrini1996,Butera1993}.
Although the values of $c_0$ and $c_1$ have rather large uncertainties,
the combination $c_0+c_1$, which we find to be $4.36(11)$, has a smaller uncertainty.
This is because the value of the quantity in Eq.~(\ref{eqn:criteria_param}) is stable 
on any line defined by $c_0+c_1= [{\rm constant}]$ 
near the minimum point. 
In Fig.~3 we plot $\tilde{a}_n$ and
$\tilde{a}^{\mbox{\tiny (K-T)}}_n$ corresponding to the best fitting values of $\beta_c$, $c_0$ and $c_1$
for $n_0=13$.
We find that the agreement between the two series 
is good not only in the range $13\le n \le 24$ used in the fitting 
but also in the range $5\le n \le 12$, which was not used in the fitting. 
This confirms the validity of our assumption that 
the series $\tilde{a}_n$ for sufficiently large $n$ is dominated 
by the Taylor expansion of the first term in Eq.~(\ref{eqn:f_cr2}).

Next, we fixed the inverse critical temperature $\beta_c$ to $1.1199$. In this case,
the fitting yields $\sigma=0.513(19)$, and $c_0+c_1=4.18(23)$.
This best fitting value of $\sigma$ is also consistent 
with the K-T value of $1/2$.

Finally, we set $\sigma$ to $1/2$ and $\beta_c$ to $1.1199$.
In this case, we obtained $c_0+c_1=4.36(11)$.
Here, the behavior of the coefficients $\tilde{a}_n$ and $\tilde{a}^{\mbox{\tiny (K-T)}}_n$
with respect to $n$ is almost the same as that plotted in Fig.3.
The fitted values of the parameters $c_0$ and $c_1$ imply 
that the non-universal constant $b$ in Eq.~(\ref{eqn:f_cr}) 
is given by $(c_0+c_1)/2\sqrt{2}=1.54(4)$.
We can see that the value of $b$ obtained from the high-temperature series 
for the free energy is consistent with the value obtained
from the Monte Carlo simulation of the correlation length
to a precision of 15\%
and with the value obtained from the high-temperature series of the correlation length
to a precision of 10\%. 

\section{Summary}

 We have calculated the high-temperature series 
for the free energy of the XY model on the square lattice
to order $\beta^{48}$ 
using an improved finite-lattice method.
This method in general enables us to generate a longer expansion series
for spin systems whose spin variable takes more than two values. 
The length of the series obtained here is two times longer than
the series obtained previously with the graphical method. 
 The results of the analysis for the obtained series 
give strong support to the validity of the prediction
given in Eq.(\ref{eqn:f_cr}) 
for the behavior of the free energy of the XY model in two dimensions.
The value of the critical point derived from the series
for the free energy is
consistent with the value obtained in studies 
of the correlation length
employing the numerical simulations and series expansions.
The value of the non-universal constant 
obtained from the series for the free energy is 
close to the value obtained from the correlation length.
 These results reconfirm  
that the phase transition of the model is 
of the Kosterlitz-Thouless type, which was first found 
in a study of the correlation length.   
\section*{Acknowledgements}
\hspace*{\parindent}
 The author would like to thank 
K. Tabata for valuable discussions. 
This work was supported in part by a Grant-in-Aid 
for Scientific Research (No.\ 16540353) 
from the Ministry of Education, Culture, Sports, Science and Technology.
%

\end{document}